\documentclass{aa}
\usepackage{graphics}
\def\dd{{\rm d}}
\newcommand{\teff}{{\ensuremath{T_{\mbox{\scriptsize eff}}}}}

\begin{document}

\thesaurus{02.04.2; 08.01.1; 08.05.3; 08.22.2}

\title{The Effect of Diffusion on Pulsations of Stars on the 
Upper Main Sequence.}
\subtitle{$\delta$~Scuti and Metallic A Stars}

\author{S. Turcotte\inst{1,2}
        \and J. Richer\inst{3} 
        \and G. Michaud\inst{3}
        \and J. Christensen-Dalsgaard\inst{2,4}}
   \offprints{S. Turcotte}
\institute{Service d'Astrophysique/DAPNIA, CEA-Saclay, Gif sur Yvette, 91191, France \\
        email: turcotte@discovery.saclay.cea.fr
        \and Teoretisk Astrofysik Center, Danmarks Grundforskningsfond
        \and Universit\'e de Montr\'eal and Centre de Recherche en Calcul Appliqu\'e, Montr\'eal, Canada \\
        email: richer@cerca.umontreal.ca and michaudg@cerca.umontreal.ca
        \and Institut for Fysik og Astronomi, Aarhus Universitet, 
	DK-8000 {\AA}rhus C, Denmark \\
        email: jcd@ifa.au.dk}

\authorrunning{Turcotte et al.}
\titlerunning{Diffusion and Pulsation in A Stars}

\date{Received ; accepted }

\maketitle

\begin{abstract}
Recent dramatic improvements in the modeling of abundance evolution
due to diffusion in A stars have been achieved with the
help of monochromatic opacity tables from the OPAL group.
An important result in the context of stellar pulsations is the
substantial helium abundance shown to be left over in the driving region of
$\delta$~Scuti-type pulsations in chemically peculiar Am stars. An accurate opacity
profile in the entire stellar envelope including the full effect of 
heavy elements is also now available for the first time.

Pulsations are shown to be excluded for young Am stars but occur naturally
when these stars evolve off the ZAMS. The predicted variable metallic A stars
all lie towards the red edge of the instability strip, in
qualitative agreement with the observed variable
$\delta$~Delphini and mild Am stars.

Results show little direct excitation from iron-peak elements in
A-type stars. The main abundance effect is due to the settling
of helium, along with a marginal effect due to the enhancement 
of hydrogen.
\end{abstract}

\section{Introduction} 

Until very recently the theory for the formation
of chemically peculiar (CP) stars
and the observations of variable stars agreed that CP stars were not variable and
variable stars were chemically normal.  A few exceptions were known for many years
but their rarity and their mild chemical anomalies did not challenge the standard
picture.

In recent years this picture has evolved considerably and the dichotomy between 
CP stars and variable stars is no longer clear-cut. 
Indeed, many types of CP stars are now known to exhibit various types
of pulsations. 
Significantly, most of the variable
CP stars belong to the magnetic Ap and Bp families. But  some non-magnetic CP stars
are known to be variable and it has been claimed that in some cases the observed variability
cannot be easily reconciled with the theory of diffusion for CP stars.

In this paper we examine how recent progress in modeling Am stars affects our
understanding of the problem of variability in A-type CP stars.  We 
first review the current observational and theoretical pictures for variability in
CP stars. The improved models of diffusion in A stars are then presented.
The linear nonadiabatic oscillation equations are solved in selected models
of A-type stars, both with and without diffusion.  
These results highlight the differential effect of diffusion on the stability 
of the models, and allow to draw conclusions on 
how well the models with diffusion can be reconciled with observations.

\section{Metallicism and pulsations: observations}\label{sec:obs}

CP stars are found almost everywhere in the HR diagram. Some of these stars
are evolved and their observed abundance peculiarities reflect nuclear processes.
Others are compact objects in which diffusion is well established 
(Michaud \& Fontaine~\cite{MichaudFontaine84}). 
Many of these objects are variable and they are very favourable for 
asteroseismology.

We concentrate mainly on main-sequence stars in which diffusion is thought to be
the principal cause of abundance anomalies. 
These are found from early F-type (Fm) stars to
late B-type (HgMn) stars, including many varieties of A stars [Am, Ap, $\lambda$~Bootis,
$\rho$~Puppis].  
Within these spectral classes many types of
variable stars are also found 
[$\gamma$~Doradus, several classes of $\delta$~Scuti, and 
Slowly Pulsating B (SPB) stars]. 

In A-type stars, almost 70\% of non-CP stars are $\delta$~Scuti
variables at current levels of sensitivity.  
Most non-variable A stars are Am stars.

Amongst the CP stars which happen to be pulsating, the most conspicuous
are the ``rapidly oscillating Ap stars" (roAp) first discovered by Kurtz~(\cite{Kurtz78a}). 
These are found amongst the coolest of the magnetic Ap stars which exhibit 
large abundance anomalies of many heavy elements. 
They have generated considerable 
interest because of the relatively large number of observed overstable modes which makes 
them promising objects for asteroseismology.
See, for example, Matthews~(\cite{Matthews91}) for a review.

In non-magnetic stars, however,
variability and anomalous abundances are found in very few
stars simultaneously.
Over the years some mildly metallic stars have been found to 
exhibit some variability. 
Baglin~(\cite{Baglin72}) suggested that if diffusion is the cause of the Am phenomenon,
Am stars should not pulsate.
Some mild Am stars and evolved CP stars were then already known to be 
variable (Kurtz~\cite{Kurtz76}) but all
classical Am stars which were thought to be variable in the early 70's were subsequently
found to be stable.
Since then Kurtz and his collaborators have been the principal
investigators in the search for variable CP stars. They have assembled a short list
of metallic stars, most if not all of them fairly evolved, which are also $\delta$~Scuti-type
pulsators (Kurtz~\cite{Kurtz78b}, \cite{Kurtz84}, \cite{Kurtz89}; Kurtz et al.~\cite{Kurtzetal95}; Kurtz~\cite{Kurtz00}).

Amongst these stars, some have been thought to be problematic. 
In particular,  Kurtz~(\cite{Kurtz89}) claimed to have discovered 
a pulsating classical Am star.
The problem lies in its apparently large abundance anomalies, not typical of
the other known  variable non-magnetic CP stars.
Also, the variable evolved Am star \object{HD40765} has been considered by 
Kurtz et al.~(\cite{Kurtzetal95}) to be problematic because of the possibly 
large surface velocities involved. 
A point that has been made repeatedly is that the large velocities caused
by pulsations, estimated in this star to be of the order of $14\,{\rm km}\,{\rm s}^{-1}$ 
at the surface, might generate turbulence in the interior which would in turn hinder
the formation of the required surface-abundance anomalies.

In addition to variable CP stars there are some $\delta$~Scuti stars that are particularly
interesting from our point of view. First there are 
some otherwise seemingly run-of-the-mill
$\delta$~Scuti stars\footnote{Amongst these is \object{20CVn} which was classified as $\delta$~Delphini by 
Kurtz~(\cite{Kurtz76})} in which peculiar abundances 
(Russell~\cite{Russell95};
Rachkovskaya~\cite{Rachkovskaya94}) 
superficially consistent with what is expected as a result of diffusion
seem to have been found.

Second, the high-amplitude $\delta$~Scuti stars are interesting 
because they are characterized 
by very small $v\sin i$ (Solano \& Fernley~\cite{SolanoFernley97}). 
As CP stars are mostly slow rotators ($v_{\rm rot} < 100\,{\rm km}\,{\rm s}^{-1}$),
one should find different velocity distributions in CP stars 
and in $\delta$~Scuti stars.
Indeed, observations show that $\delta$~Scuti stars are on the average fairly 
fast-rotating stars, with the exception of the high-amplitude stars.
In other respects the high-amplitude $\delta$~Scuti stars
do not differ from their
more common low-amplitude counterparts.
They are, however, known to be evolved stars
and as such might not feature significant abundance anomalies.

Although observations do not completely rule out variability in Am stars,
they do pose rigorous constraints on them. Either it is an extremely rare
occurrence or pulsations are of extremely low amplitude.
An extensive search of the Hipparcos database by many authors revealed many
new variable stars (Aerts et al.~\cite{Aertsetal98}; Waelkens et al.~\cite{Waelkensetal98};
Paunzen \& Maitzen~\cite{PaunzenMaitzen98}). 
Significantly, all the newly discovered variable CP stars were
found to be magnetic (Ap or Bp). 
All the other variable stars found in this way
are of the known families of variable stars, 
{\sl i.e.}, $\gamma$~Doradus, SPB and a few $\beta$~Cephei stars.
The expected sensitivity of these surveys ranges
from as low as 3 to 55 mmag depending on the 
brightness of the object (Eyer \& Grenon~\cite{EyerGrenon98}).

The CP stars found most recently to be variable are $\lambda$~Bootis stars
in which nonradial pulsations have been detected 
(Paunzen et al.~\cite{Paunzenetal98}). 

\section{Metallicism and pulsations: theory}\label{sec:theo}

Elements migrate with respect to each other because of differential forces mostly due to
inward gravity and outward radiative pressure. This segregation of different
atomic species is what is termed here diffusion.  Diffusion is a rather fragile process because
typical diffusion velocities are of the order of fractions of cm per second.  
Large-scale motion of matter or turbulence quickly 
overwhelm diffusion and homogenize the chemical
composition outside of nuclear burning regions.
Diffusion is efficient only in stars where the competing processes are weak.
This explains why CP stars are mainly of spectral types
ranging from early F  to late B-type stars. In these spectral types the surface convection zone
is thin enough and the mass loss rates small enough (a mass loss of 
around $10^{-14}$ M$_{\sun}\,{\rm yr}^{-1}$ is large enough to remove 
all anomalies)
to permit the development of significant abundance anomalies.
As a result, CP stars are typically relatively unevolved,
slowly rotating stars. 
There are some CP stars, Ap or $\lambda$~Bootis stars, for example,
in which other factors come into play.

In the standard picture for FmAm, HgMn and Ap stars, the observed abundance anomalies of 
heavy elements develop as a consequence of the settling of helium.  
As it disappears from the superficial convection zone
it can no longer provide the opacity to sustain convection in the \ion{He}{II} zone and
the result is a much thinner convection zone due to the ionization of \ion{H}{I}. 
At this depth, the radiative forces on the various heavy elements are compatible with the pattern 
of surface-abundance anomalies. 
This model agrees qualitatively with observations but requires 
additional assumptions to obtain a quantitative agreement as the predicted anomalies are
generally too large. No calculation based on this model has ever reproduced the abundances of
individual Am stars.

The major consequence of this model
for stellar stability is that helium is no longer present to excite pulsations typical of the
classical instability strip. 
Many studies related to this problem have been carried out.
The most thorough, by Cox et al.~(\cite{Coxetal79}), 
showed that variability is possible with a low helium content 
but that the width of the instability strip decreases as the helium abundance decreases.
Their conclusion was that classical Am stars should not be variable  but that metallicism
and variability were not mutually exclusive in the red part
of the classical instability strip if
the surface helium abundance fell marginally below 0.1 but without dropping to a very low
value.

When the diffusion of heavy elements is included in a consistent fashion this picture changes
drastically. These models are dubbed here The New Montreal Models (NMM).

Magnetic Ap stars and $\lambda$~Bootis stars will be ignored at this time. 
We note in passing that for roAp stars a few models 
have been proposed based on the so-called $\kappa$ mechanism,
where the dominant driving comes from the effects of the opacity $\kappa$.
As only microscopic diffusion can explain the observed abundances at present,
the proposed models include its effects.
Amongst the possibilities are: the replenishment of 
superficial helium by advection from mass loss (Vauclair \& Dolez~\cite{VauclairDolez90}), 
the replacement of helium by  silicon pushed in the driving region by radiative levitation
(Matthews~\cite{Matthews88}), and hydrogen overabundances in the \ion{H}{I} ionization
zone as a result of helium settling (Dziembowski \& Goode~\cite{DziembowskiGoode96}).  
As for $\lambda$~Bootis stars, the currently preferred but so far unproven
accretion models for these stars 
assume that the helium abundance would remain normal in the driving region 
(Turcotte \& Charbonneau~\cite{TurcotteCharbonneau93}), accounting for 
the necessary opacity for the $\kappa$ mechanism to work. 

\section{The New Montreal Models} \label{sec:NMM}

\subsection{New results with the NMM}

The NMM include the diffusion of all major elements up to nickel consistently by 
using monochromatic opacity tables of the Livermore group (Iglesias \& Roger~\cite{OPAL96}) 
to compute the
opacity and the radiative forces accurately at all points in the star 
and for the local chemical composition;
the basic procedures were outlined by Turcotte et al.~(\cite{TRMIR98})
and Richer et al.~(\cite{Richeretal98}).

The evolution 
of the abundances and of the structure 
is completely consistent through the opacity. 
One very important property of these models is the presence of a convectively
unstable zone around $200\,000$~K where iron-group elements 
dominate the opacity.
This convection zone appears naturally as the consequence of abundance changes
if they are large enough.
The NMM then assume that this deeper convection zone is connected to the helium and 
hydrogen convection zones through convective overshoot.

The large depth of the convective mixing relative to standard models for chemically peculiar
stars results in much smaller surface-abundance variations, more in line with
observations. 
Richer, Michaud \& Turcotte~(\cite{RMT00}) showed that the radiative forces at the base of 
the iron convection zone 
follow the correct pattern over a relatively narrow region for the
Am signature to be recovered without additional assumptions. A quantitative agreement with
observed abundances for Am stars does require additional turbulence below the 
superficial convective envelope.
In the NMM, the mixing necessary to reproduce the abundances observed in Am stars prevents
the formation of this convection zone.

A significant result in the context of stellar pulsations is that 
helium is still substantially present in the \ion{He}{II} driving region in these new models.
Also, as the opacity in the vicinity of the so-called 
``metal opacity bump''\footnote{
We shall refrain from using this expression and otherwise use the expression 
``iron-peak-element opacity bump'',
as those elements dominate the opacity there}
is increased  relative to standard models, one can expect that this region will contribute to 
the excitation of longer-period modes. 
On the other hand, in some extreme cases this region might be 
convective in contrast to standard models, 
which would reduce the radiative flux and 
by consequence the driving in that region.

We shall examine the consequence of these results 
on the possible instability of Am stars. 
As a complement, we shall also estimate the effect of mild abundance 
anomalies on predicted pulsations of $\delta$~Scuti stars. 

\subsection{The basic properties of the NMM}

The NMM include the detailed diffusion of 21 major elements from H through Ni plus 
several light elements and isotopes for a total of 28 species.
The opacity data used in the evolution code and in the following analysis are
the OPAL monochromatic opacity tables (Iglesias \& Rogers~\cite{OPAL96}) which 
allow us to calculate accurate Rosseland mean opacities and radiative forces 
for any peculiar chemical composition necessary.
At low temperatures, for which the OPAL data is lacking, we supplement them with 
the Kurucz~(\cite{Kurucz91}) opacity tables. 
Although it would be desirable to take into account changes in the composition
of individual heavy elements in the atmosphere, the transition occurs at such low temperatures
that it might be of little consequence for the stellar interior.

The models incorporate standard procedures for the equation of state and the nuclear
reaction rates. 
The mixing-length formalism for convection is used and is calibrated
using the Sun (Turcotte et al.~\cite{TRMIR98}). All models have an identical, homogeneous, 
initial chemical
composition as specified by Turcotte et al.~(\cite{TRM98}).
All models are one-dimensional,
non-rotating, and non-magnetic.

Individual models for a given stellar mass
only differ in the parameters adopted for the coefficient of turbulent
diffusion. 
In all calculations presented in this paper, the coefficients are chosen so that the 
zone mixed by turbulence goes from the surface 
to somewhat below the iron convection zone.
The coefficient of turbulent diffusion is modeled 
with the following three-parameter expression
\begin{equation}
   D_{\rm T}= D_0 \left({\rho_0\over\rho}\right)^n \; ,
\label{eq:Dt}
\end{equation}
where the free parameters are $D_0$, $\rho_0$ and $n$. 
The evolution of the abundances
is very sensitive to the depth of the mixing 
but not so much to the profile of $D_{\rm T}$.
The models are named with reference to the number $R$ 
which specifies the ratio of $D_{\rm T}$
to the coefficient of atomic diffusion of helium $D_{\rm He}$
at the point where the density $\rho$ is
equal to the reference value $\rho_0$. 
For example, model 1.90R1000-2 is a 1.90~M$_{\sun}$ star 
with $R=1000$ and $n=2$;
for simplicity, `10K' is used to refer to models with $R = 10\,000$.
In all the models discussed in this paper
$\rho_0$ is $8\times 10^{-6}$g\,cm$^{-3}$. 
The reader is referred to Richer et al.~(\cite{RMT00})
and Richard et al.~(\cite{RMR00}) for further details.

For every stellar mass examined, 
one model that does not include any effects of diffusion is also included. 
For computational efficiency this comparison model uses the mean Rosseland opacity 
tables of Iglesias \& Rogers~(\cite{OPAL96}). Assuming that there is no separation
of elements in the star implies that some unspecified mixing is necessarily assumed. This
mixing is required to be large and deep enough to keep superficial regions at a 
constant chemical composition without mixing too deeply, to avoid
dredging up nuclearly processed matter to the surface.
They are named according to the mass and are labeled with
the tag ``ND'' ({\sl e.g.}, 1.90-ND).

\section{Diffusion and the $\kappa$ mechanism}\label{sec:diffusion}

To determine the frequencies of modes of oscillation for a star requires only
that we solve the adiabatic equations.
Solving the full nonadiabatic equations of stellar oscillation
allows us to calculate the growth rates of the modes, and
hence to determine 
which of the modes are overstable;
also, by considering the work integral we can investigate the
contributions of the different parts of the star to the excitation and
damping of the mode.

The nonadiabatic oscillation package used was generously provided to us by 
W. Dziembowski and follows the procedure first described by
Dziembowski~(\cite{Dziembowski77}). We are mainly
concerned here with excitation via the $\kappa$ mechanism on which abundance
variations have a direct impact. 
We note,
however, that the present calculations lack a good modeling of 
the effect of convection;
this must be kept in mind in the analysis of the results.

The physics of the $\kappa$ mechanism has been reviewed extensively 
({\sl e.g.} Cox~\cite{Cox80}; Unno et al.~\cite{Unnoetal79}; 
Gautschy \& Saio~\cite{GautschySaio95}).
As a quick reminder, generating pulsations in a star requires that the energy gained 
by an oscillation mode over a complete cycle be larger than the energy lost.
We are then looking for a positive net work over the entire star over one cycle. 
In the case of the $\kappa$ mechanism, the energy is transferred from the 
outward radiation flux to the oscillation mode via the opacity. 
A mode becomes overstable
by this mechanism if the opacity profile and its derivatives have the right
features.

Following Unno et al.~(\cite{Unnoetal79}), 
from the definition of the work ($W$) as
the variation of the kinetic energy ($E$) over a cycle,
\begin{equation}
  W=\oint { \dd E\over \dd t}  \dd t \, ,
\end{equation}
one can write
\begin{equation}
  W={\pi\over\omega}\int_0^{M_r} {\delta T\over T}
\delta\left[\epsilon_{\rm N}-{1\over\rho}\nabla\dot
    (F_{\rm R}+F_{\rm c})\right] \dd M_r \, ,
\end{equation}
where $\delta$ denotes Lagrangian perturbations.
Also, $\omega$ is the (angular) oscillation frequency,
$T$ is temperature, $M_r$ is the mass interior to the radius $r$,
$\epsilon_{\rm N}$ is the nuclear energy generation rate,
and $F_{\rm R}$ and $F_{\rm c}$ are the radiative and convective
fluxes.
If one neglects the contribution from the nuclear ($\epsilon_{\rm N}$)
and convective terms ($F_{\rm c}$),
and only keeps the perturbation of the radiative flux ($F_{\rm R}$),
one can isolate the
contribution of the $\kappa$ mechanism 
to the driving of a given mode of oscillation.
To obtain a simple estimate of
this contribution to
the work integral we make the quasi-adiabatic approximation
({\sl i.e.}, evaluate the work integral by means 
of adiabatic eigenfunctions),
and furthermore assume that the 
adiabatic thermodynamic derivatives $\Gamma_1$,
$\Gamma_3$ and $\nabla_{\rm ad}$ are constant.
Then the work done by the $\kappa$ mechanism is proportional to
\begin{equation}
   \int \left({\delta T\over T}\right)^2 {\dd \over \dd r}
\left[\left(\kappa_T +
          {\kappa_\rho\over\Gamma_3-1}\right)L_r\right] \dd M_r \; ,
\end{equation}
where $L_r$ is the luminosity at $r$,
and $\kappa_T=(\partial\ln\kappa/\partial\ln T)_\rho$,
$\kappa_\rho=(\partial\ln\kappa/\partial\ln\rho)_T$;
thus regions where
\begin{equation}
   {\dd \over \dd r}\left(\kappa_T + {\kappa_\rho\over\Gamma_3-1}\right) > 0
\label{eq:dW}
\end{equation}
contribute to the excitation. 
Local increases in the logarithmic derivatives of $\kappa$ are
necessary and  a decrease in $\Gamma_3-1$ in partial ionization zones of
a dominant species (H or He) is helpful.  It also follows that regions where the gradients of
$\kappa_T$ and $\kappa_\rho$ are negative contribute to 
damping of the pulsation.
The $\kappa_T$ term usually dominates over the other term.

The numerical results reported in the text, 
including the growth parameters and work integrals, are
computed using the full nonadiabatic procedure of the Dziembowski code.

In order that the excitation by the $\kappa$ mechanism should not 
be cancelled by damping elsewhere, it is necessary that the
driving region lie in the so-called
transition zone between the quasi-adiabatic and nonadiabatic regimes;
in that case, the oscillations are strongly nonadiabatic outside
the driving region, and this part of the star therefore does not
contribute to the damping, giving rise to net driving.
This leads to an approximate relation between the period $\Pi$
of a given mode of pulsation and the position of the transition region in a
star (Cox~\cite{Cox80}):
\begin{equation}
   {\langle c_v T\rangle_{\rm tr} \Delta M_{\rm tr}\over L \Pi} \simeq 1 \, ,
   \label{eq:period}
\end{equation}
here $\Delta M_{\rm tr}$
is the mass outside the transition region,
$\langle \cdots \rangle_{\rm tr}$ is the average over that part of the
star, $c_v$ being the specific heat, and $L$ is the luminosity.

The normalized growth rate is defined as 
\begin{equation}
  \eta = \int {\dd W\over \dd \log T} \dd \log T \left/ 
\int \left|{\dd W\over \dd \log T}\right| \dd \log T \right. \, .
\end{equation}
In this formulation, $\eta$ varies from $+1$,
if there is driving in the entire star,
to $-1$, if there is damping in the entire star.
The value of zero defines neutral stability.

Diffusion affects the $\kappa$ mechanism by decreasing
driving from helium in favour of driving from metals.
As a consequence of Eq.~(\ref{eq:period}),
the pulsation period of the unstable modes
depends on the depth of the driving region. 
During a star's evolution the helium ionization zone gradually shifts deeper in the star,
thereby increasing the period of the observed pulsation modes.
Additionally, as the driving in the deeper iron-peak driving region increases
while the driving due to helium  decreases as a result of diffusion, 
one might expect the observed pulsation periods to shift to even longer periods.
The effect of abundance variations on the opacity profiles is discussed 
below for selected models (see Fig.~\ref{fig:A1.9opac}).

\section{Opacity and its derivatives} \label{sec:opacity}

We recomputed many thermodynamic quantities and the opacities in the process
of preparing the models for input to the nonadiabatic oscillation package.
Every effort has been made to be consistent with the procedures followed in the
evolution code. The most tricky operation at this point is the determination
of accurate opacity derivatives.
The opacities are interpolated
linearly and smoothed locally in the evolution code.  As first and second derivatives of the
opacity are needed in the oscillation package a more refined
interpolation procedure had to be adopted to guarantee smoother derivatives.
We chose the Houdek (Houdek \& Rogl~\cite{HoudekRogl96}) routines which use 
two-dimensional rational splines.
For each mesh point of the models we constructed a $7\times 7$ opacity grid
in the $(\rho,T)$ plane in which the splines were fitted and the
opacity derivatives were determined.

\section{The models of A stars} \label{sec:Astars}

Amongst the numerous NMM models some were selected to provide a 
satisfactory sampling of the instability strip and of Am stars. 
The selected models of 1.9, 2.0 and 2.2~M$_{\sun}$ are summarized in 
Tables~\ref{tab:1.9models}, \ref{tab:2.0models}, 
and~\ref{tab:2.2models} respectively.
(The naming convention for the models was described in the 
discussion relating to Eq.~\ref{eq:Dt}.)
Of these, the models with $R=1000$ and
$n=2$ reproduce reasonably well the abundances of Am stars 
(Richer et al.~\cite{RMT00}). 
They are not optimal, however,
and models with $R=1000$ and with shallower turbulence, $n=3$ or 4, 
would be more accurate. 
We feel that using such models would not alter the essence of the interpretation of the results.

\begin{table}
\begin{center}
\begin{tabular}{rcccc}
\hline
\hline
  Age   & $L/L_{\sun}$  & \teff & $R/R_{\sun}$  & $\log g$       \\
  (Myr) & \strut        & \strut        & \strut        & \strut        \\
\hline
\multispan 5  \hfill\strut 1.90-ND  \hfill \\
\hline
 101    & 13.50         & 8710          & 1.628         & 4.293  \\
 300    & 14.30         & 8501          & 1.759         & 4.226  \\
 502    & 15.17         & 8248          & 1.924         & 4.148  \\
 670    & 15.94         & 7952          & 2.123         & 4.063  \\
 1\,003 & 17.04         & 7002          & 2.830         & 3.813  \\
\hline
\multispan 5  \hfill\strut 1.90R10K-2  \hfill \\
\hline
 101    & 13.09         & 8669          & 1.615         &  4.291 \\
 300    & 14.11         & 8431          & 1.633         &  4.217 \\
 503    & 15.01         & 8148          & 1.961         &  4.131 \\
 670    & 15.76         & 7848          & 2.167         &  4.045 \\
 1\,002 & 16.84         & 6919          & 2.881         &  3.797 \\
 1\,105 & 23.72         & 7063          & 3.281         &  3.684 \\
\hline
\multispan 5  \hfill\strut 1.90R1000-2  \hfill \\
\hline
 101    & 13.34         & 8652          & 1.640         & 4.287  \\
 300    & 14.11         & 8407          & 1.787         & 4.212  \\
 503    & 15.04         & 8125          & 1.975         & 4.125  \\
 670    & 15.82         & 7830          & 2.181         & 4.039  \\
 1\,001 & 16.96         & 6918          & 2.893         & 3.794  \\
 1\,094 & 19.29         & 7052          & 2.969         & 3.771  \\
\hline
\multispan 5  \hfill\strut 1.90R300-2  \hfill \\
\hline
 101    & 13.34         & 8640          & 1.645         & 4.284  \\
 300    & 14.11         & 8387          & 1.795         & 4.208  \\
 503    & 15.04         & 8103          & 1.985         & 4.121  \\
 670    & 15.82         & 7809          & 2.192         & 4.035  \\
 1\,003 & 16.98         & 6898          & 2.912         & 3.788  \\
 1\,098 & 19.00         & 6958          & 3.027         & 3.755  \\
\hline
\hline
\end{tabular}
\end{center}
\caption{Physical parameters of the 1.9~M$_{\sun}$ models}
\label{tab:1.9models}
 \end{table}

\begin{table}
\begin{center}
\begin{tabular}{ccccc}
\hline
\hline
  Age   & $L/L_{\sun}$  & \teff & $R/R_{\sun}$  & $\log g$       \\
  (Myr) & \strut        & \strut        & \strut        & \strut        \\
\hline
\multispan 5  \hfill\strut 2.00-ND  \hfill \\
\hline
 300   & 17.84          & 8767          & 1.847         & 4.206  \\
 670   & 20.27          & 7979          & 2.377         & 3.987  \\
 801   & 20.85          & 7501          & 2.728         & 3.867  \\
 852   & 21.06          & 7290          & 2.903         & 3.813  \\
\hline
\multispan 5  \hfill\strut 2.00R2500-0.75  \hfill \\
\hline
 300   & 17.62          & 8731          & 1.851         & 4.204  \\
 670   & 20.06          & 7958          & 2.378         & 3.986  \\
 801   & 20.69          & 7497          & 2.720         & 3.870  \\
 854   & 20.88          & 7274          & 2.903         & 3.813  \\
\hline
\multispan 5  \hfill\strut 2.00R1000-2  \hfill \\
\hline
 300   & 17.59          & 8663          & 1.879         & 4.191  \\
 670   & 19.97          & 7862          & 2.430         & 3.967  \\
 751   & 20.37          & 7594          & 2.631         & 3.898  \\
 854   & 20.75          & 7186          & 2.966         & 3.794  \\
\hline
\hline
\end{tabular}
\end{center}
\caption{Physical parameters of the 2.0~M$_{\sun}$ models}
\label{tab:2.0models}
 \end{table}

\begin{table}
\begin{center}
\begin{tabular}{ccccc}
\hline
\hline
  Age   & $L/L_{\sun}$  & \teff & $R/R_{\sun}$  & $\log g$       \\
  (Myr) & \strut        & \strut        & \strut        & \strut        \\
\hline
\multispan 5  \hfill\strut 2.20-ND  \hfill \\
\hline
 500   & 29.88          & 8595          & 2.487         & 3.989 \\
 603   & 30.97          & 8082          & 2.864         & 3.866 \\
 670   & 31.71          & 7700          & 3.193         & 3.772 \\
\hline
\multispan 5  \hfill\strut 2.20R1000-2  \hfill \\
\hline
 300   & 26.78          & 9116          & 2.093         & 4.138 \\
 501   & 29.48          & 8470          & 2.545         & 3.969 \\
 597   & 30.52          & 8001          & 2.901         & 3.855 \\
\hline
\hline
\end{tabular}
\end{center}
\caption{Physical parameters of the 2.2~M$_{\sun}$ models}
\label{tab:2.2models}
 \end{table}
   \begin{figure}
      \resizebox{\hsize}{!}{\includegraphics{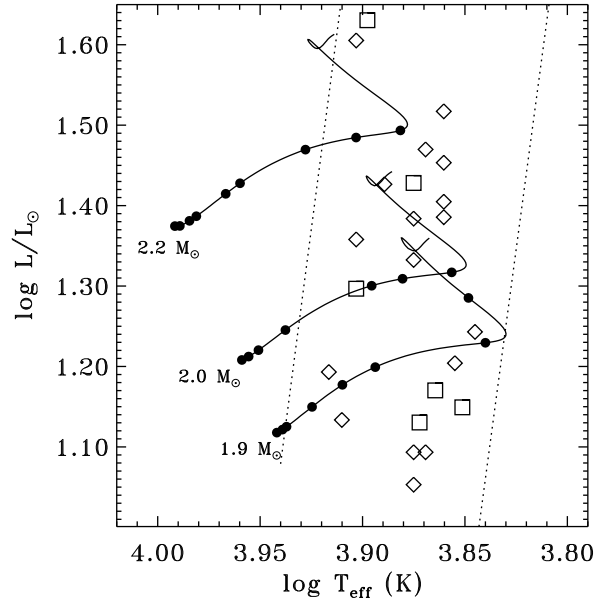}}
      \caption[]{Various classes of peculiar and variable A stars in the 
                 HR diagram and evolutionary paths for some of our models 
                 whose mass are indicated in the graph 
                 (these paths are for models R1000-2; paths for other models
                 at the same masses would be indistinguishable at this scale).
                 The squares identify
                 the stars of Kurtz~(\cite{Kurtz76}) while the diamonds identify the
                 $\delta$~Scuti stars observed by Russell~(\cite{Russell95}).
                 The black dots mark the position of the models listed in Tables~\ref{tab:1.9models}
                 through \ref{tab:2.2models}. The dotted lines
                 identify the approximate boundaries of the instability strip.
              }
         \label{fig:AHRD}
   \end{figure}
%

\section{$\delta$~Scuti-type pulsations in Am stars}\label{sec:drive} 

Most pulsating A stars lie in the \ion{He}{II} instability
strip. Consequently, most attempts to model the pulsations of chemically
peculiar A stars have centered on how helium can remain or be replenished
in the \ion{He}{II} zone, or possibly which element can replace it as
the motor of the excitation mechanism. 
As previously discussed, an important characteristic of the NMM is the 
relatively high superficial helium abundance remaining throughout the evolution
in Am stars (see Sect.~\ref{sec:NMM}).
This section aims at answering some questions raised by these new models:
Is helium sufficiently abundant to generate variability in Am stars?
Do the NMM clash with our current observational understanding of metallicism and
pulsations? Does the larger driving in the iron-peak opacity bump play a role
in exciting g modes in CP stars?

\subsection{The variability of an Am star across the instability strip}
\label{sec:Am}

The evolution of a 1.9~M$_{\sun}$ star from the ZAMS to hydrogen exhaustion in the
core spans almost the full width of the instability strip as shown in 
Fig.~\ref{fig:AHRD}. 
Studying the evolution of instability in models for such a star as it evolves
gives a first idea of the behaviour of Am stars in general.
The luminosity dependence of the instability strip implies that the 
lower part of the instability strip will be populated by 
young stars and the upper part by generally more evolved 
stars. This should be taken into account if one attempts
to extrapolate the following results to stars of different 
masses. 

Tables~\ref{tab:overstable} and~\ref{tab:overstable1} show the general pattern of
overstability for radial ($l=0$) and nonradial ($l=1$) p modes, respectively,
in an evolving 1.9~M$_{\sun}$ model with and without diffusion. The model
with diffusion shown here is an adequate representation of a classical Am star
(see Fig. 22 of Richer et al.~\cite{RMT00}).

First, one can see that no overstable modes, radial or 
otherwise, are found in the young model with diffusion (Am star) 
whereas high-order modes are predicted to be overstable in the model without 
diffusion (a standard model for  $\delta$~Scuti stars, {\sl i.e.},
a homogeneous envelope  with solar composition). 
As the stars age, the lower-order modes are gradually more excited and eventually 
the fundamental mode becomes overstable in both models. In the model with
diffusion, the superficial helium abundance reaches its lowest value at 750 Myr 
where it is 0.114 by mass. The model is stable 
at that time. The more evolved models tend to be more unstable. At an age
of 1.098 Gyr, for example, the helium mass fraction rises to 0.126. The 
small difference in the helium abundance for these two
models suggests that the pulsations in evolved stars occur 
without relying on the dredge-up of helium nor on 
hypothetic mass-loss but only as a result of their evolution. 
This result is in agreement with Cox et al.~(\cite{Coxetal79}) and
earlier work which showed that stars would be expected to be variable in the
red part of the instability strip if their helium abundance were of the order of 0.1
or higher.

As far as nonradial modes are concerned, the same overall behaviour as for
radial modes is repeated,
although the labeling of the modes presents some problems.
Nonetheless, a presentation such as Table~\ref{tab:overstable1} 
illustrates the general evolution of overstability in nonradial modes;
here
the mathematical mode orders, defined such that the mode order of a given mode
is constant during the evolution of the model, are shown.
This is not directly related to the physical nature of the mode, however.
For p modes, including radial modes, the dimensionless frequency%
\footnote{e.g. $\sigma$, defined by $\sigma^2 = R^3 \omega^2/(G M)$
where $M$ and $R$ are total mass and surface radius of the star,
and $G$ is the gravitational constant.},
normalized with the dynamical timescale of the star,
is approximately constant during the evolution.
In contrast, the dimensionless frequency increases for g modes,
which are strongly affected by the composition structure in the stellar
interior.
Where the frequency of a g modes meets the frequency of a p mode,
an avoided crossing takes place,
at which occurs a shift 
in mode order for a given frequency, or conversely a shift in frequency 
for a given mode order (e.g.\ Osaki 1975).
Thus, in contrast to radial modes, there is no firm correspondence
between mode order and dimensionless frequency through the life of a star.
Even so, a table corresponding to Table~\ref{tab:overstable1} but based
on a definition of the order more closely related to the physical nature
of the modes, would have shown a similar overall pattern.
The more evolved models feature overstable nonradial p and
g modes. In the most evolved model, the last model of 1.90-ND, 
the overstable modes all have the physical nature of g modes,
spanning periods of 1 to 3.5 hours. Some modes, at avoided
crossings, are mixed modes which share p and g mode characters.

Two models with a turbulence depth bracketing that of the model discussed so far
are also available at this mass.
In the model with shallower mixing (1.90R300-2) 
only one mode is found to have a positive growth rate, the $l=0,n=3$ mode for 
the model at 1.098~Gyr. No overstable nonradial modes have been found. The
minimum helium abundance at 750~Myr is 0.0734 
and increases to 0.0871 at 1.098~Gyr for this star.
In the model with deeper mixing (1.90R10K-2) the minimum helium abundance reached 
is higher (0.180) and the youngest age at which overstability is found is 850~Myr.
This confirms that the blue edge of the instability strip 
is sensitive to the helium abundance,
which has long been established (Cox et al.~\cite{Coxetal79} and references therein).

\begin{table}
\begin{center}
\begin{tabular}{r|c|c|c|c|c|c|c|c}
\hline
\hline
\multispan 9 \hfill\strut $l=0$ \hfill \\
\hline
\strut    & \multispan 8 \hfill\strut $n$ \hfill \\
age (Myr) & 1 & 2 & 3 & 4 & 5 & 6 & 7 & 8 \\
\hline
\multispan 9 \hfill\strut 1.90R1000-2 \hfill \\
\hline
 101      &   &   &   &   &   &   &   &\strut  \\
 300      &   &   &   &   &   &   &   &\strut  \\
 503      &   &   &   &   &   &   &   &\strut  \\
 670      &   &   &   &   &   &   &   &\strut  \\
 1\,003   &$+$&$+$&$+$&$+$&$+$&   &   &\strut  \\
 1\,098   &$+$&$+$&$+$&$+$&   &   &   &\strut  \\
\hline
\multispan 9 \hfill\strut 1.90-ND \hfill \\
\hline
 101      &   &   &   &   &   &   &   &$+$\strut\\
 300      &   &   &   &   &   &   &$+$&$+$\strut\\
 502      &   &   &   &   &   &$+$&$+$&   \strut\\
 670      &   &   &   &$+$&$+$&$+$&$+$&   \strut\\
 1\,003   &$+$&$+$&$+$&$+$&$+$&$+$&$+$&   \strut\\
 1\,105   &$+$&$+$&$+$&$+$&$+$&$+$&   &   \strut\\
\hline
\hline
\end{tabular}
\end{center}
\caption{Overstable radial modes in models 1.90R1000-2 and 1.90-ND for several ages.
         $+$ identifies models for which there is at least one mode with $\eta>0.0$.}
\label{tab:overstable}
 \end{table}

\begin{table}
\begin{center}
\begin{tabular}{r|c|c|c|c|c|c|c|c}
\hline
\hline
\multispan 9 \hfill\strut $l=1$ \hfill \\
\hline
\strut    & \multispan 8 \hfill\strut $n$ \hfill \\
age (Myr) &        1 & 2 & 3 & 4 & 5 & 6 & 7 & 8 \\
\hline
\multispan 9 \hfill\strut 1.90R1000-2 \hfill \\
\hline
 101      &\strut    &   &   &   &   &   &   &   \\
 300      &\strut    &   &   &   &   &   &   &   \\
 503      &\strut    &   &   &   &   &   &   &   \\
 670      &\strut    &   &   &   &   &   &   &   \\
 1\,003   &\strut $+$&$+$&$+$&$+$&   &   &   &   \\
 1\,098   &\strut $+$&$+$&$+$&   &   &   &   &   \\
\hline
\multispan 9 \hfill\strut 1.90-ND \hfill \\
\hline
 101      &\strut    &   &   &   &   &$+$&$+$&$+$\\
 300      &\strut    &   &   &   &   &$+$&$+$&   \\
 502      &\strut    &   &   &   &$+$&$+$&$+$&   \\
 670      &\strut    &$+$&$+$&$+$&$+$&$+$&   &   \\
 1\,003   &\strut $+$&$+$&$+$&$+$&$+$&   &   &   \\
 1\,105   &\strut    &   &   &   &   &   &   &   \\
\hline
\hline
\end{tabular}
\end{center}
\caption{Overstable nonradial ($l=1$) p modes in models 1.90R1000-2 and 1.90-ND for several ages.
         $+$ identifies models for which there is at least one mode with $\eta>0.0$.}
\label{tab:overstable1}
 \end{table}

\begin{figure}
      \resizebox{\hsize}{!}{\includegraphics{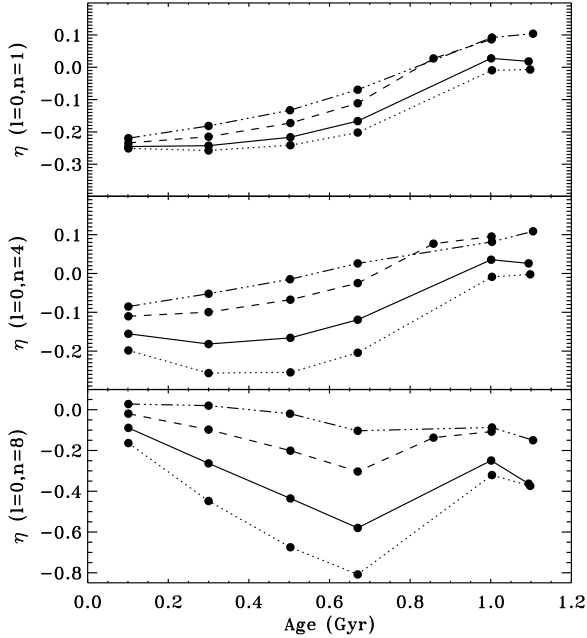}}
      \caption[]{Variation of the growth rate of selected radial modes as a function of age
                 in models of 1.90 M$_{\sun}$. 
		 The dot-dashed line represents model 1.90-ND,
		 the dashed line 1.90R10K-2,
                 the solid line 1.90R1000-2,
		 and the dotted line 1.90R300-2.
              }
         \label{fig:A1.9eta}
   \end{figure}

In an attempt to give a better picture of the evolution
of the excitation of various
modes in the star, Fig.~\ref{fig:A1.9eta} shows the change in the growth rate as a function
of time for 3 different radial modes in all four models
of the 1.9~M$_{\sun}$ star.
This figure reflects the preceding discussion in the sense that the growth rates of low-order
modes increase with passing time and that high-order modes become less excited.
It also shows that the effect of diffusion is much stronger for higher-order modes.
The different evolution of the growth rates of the $l=0,n=8$ mode after 700~Myr
in the models with and without diffusion is noteworthy. This will be discussed again
shortly.

%

\begin{figure}
      \resizebox{\hsize}{!}{\includegraphics{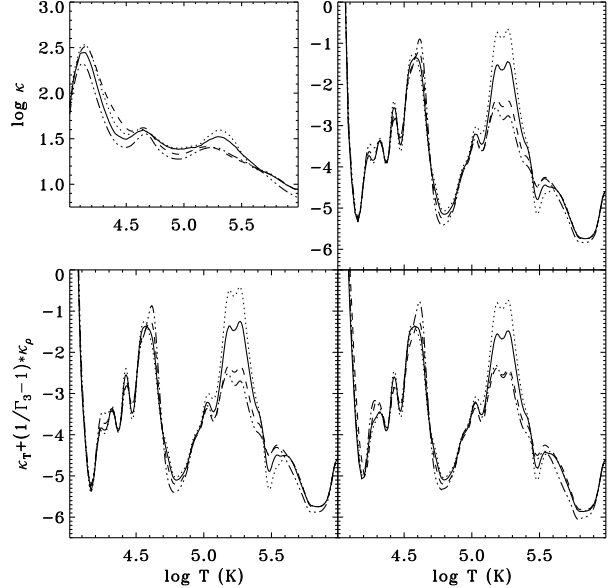}}
      \caption[]{Evolution of the opacity in the various 1.90 M$_\odot$ models discussed
                 in the text. The panel on the upper left shows the logarithm of
                 the Rosseland mean opacity at an age of 670~Myr for all models. The
                 other three panels show the variation of the argument 
		 of Eq.~(\ref{eq:dW})
                 at three ages: upper right at 100~Myr, lower left at 670~Myr, and
                 lower right at 1000~Myr. The curves are associated with
		 the following models:
		 dot-dashed line for 1.90-ND,
		 dashed line for 1.90R10K-2,
                 solid line for 1.90R1000-2,
		 and dotted line for 1.90R300-2.
                 }
         \label{fig:A1.9opac}
   \end{figure}

The detailed effect of diffusion on the opacities of A stars is
illustrated in Fig.~\ref{fig:A1.9opac} where the opacity is shown for the same four models
at 670~Myr.  The other panels show the evolution of the argument of Eq.~(\ref{eq:dW})
for ages of roughly 100, 670 and 1000~Myr.  
The most striking effect is
the increase around $\log T\simeq 5.3$ caused 
by the heightened iron-peak opacity bump. 
The effect of the settling of helium around $\log T\simeq 4.5$ appears
minor in the figure, but this ends up as being the dominant effect.  

\begin{figure*}
      \resizebox{\hsize}{!}{\includegraphics{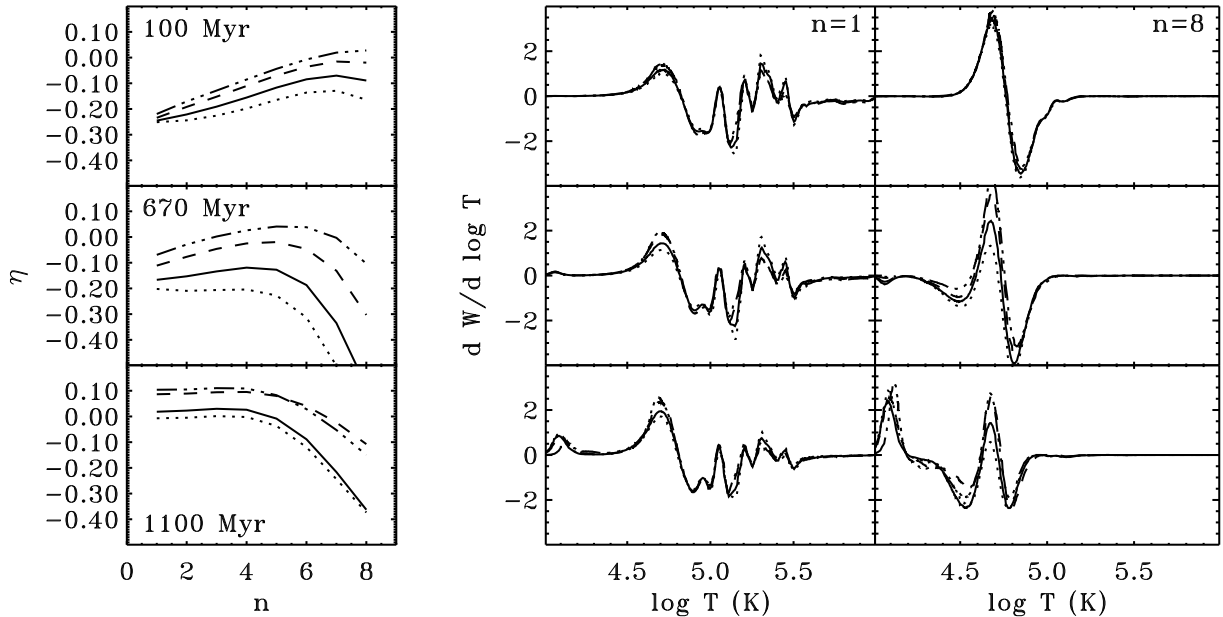}}
      \caption[]{Growth rates ($\eta$) for low-order radial modes $l=0$ (left) and the work integrals
                 for the fundamental $l=0,n=1$ (centre) and the shorter period $n=8$ (right)
                 modes for models of 1.9~M$_{\sun}$
                 at 100, 670, and 1100~Myr from top to bottom respectively.
                 The dot-dashed line represents model 1.90-ND,
		 the dashed line 1.90R10K-2,
		 the solid line 1.90R1000-2, 
                 and the dotted line 1.90R300-2.
              }
         \label{fig:Aradial1.9}
   \end{figure*}

Considering just Fig.~\ref{fig:A1.9opac} can give a false impression
of the importance of diffusion in different regions.
A truer impression is obtained from
Fig.~\ref{fig:Aradial1.9} where the growth rates for radial modes are shown
together with the work integrals for two of those modes, the $l=0,n=1$ and the
$l=0,n=8$, again at 100, 670 and 1000~Myr. 

Evidently, the work integrals for different models are close to each other in
most regions of the star. The differences are larger for the $n=8$ modes which is reflected
in the growth rates. The driving due to helium at $\log T\simeq 4.7$ is the feature
most obviously affected by diffusion, decreasing as helium settles out of the
convection zone. Diffusion has some marginal effect in the iron-peak 
opacity bump for the $n=1$ mode where it partially compensates the effect of He
settling but it has no effect whatsoever for the higher-frequency $n=8$ mode. 
The differences from model to model are more easily seen in the growth rates which have already been discussed. 

Comparing the evolution of driving in the models with and without diffusion
in Fig.~\ref{fig:Aradial1.9} explains the different evolution of the growth rates
of low and high-order modes in the presence of diffusion
presented in Fig.~\ref{fig:A1.9eta}. In Fig.~\ref{fig:Aradial1.9},
the height of the peaks due to \ion{H}{I}, at $\log T\simeq 4.1$,
increases with time 
at a rate quite similar in all models shown. However,
the evolution of the peak due to helium, at $\log T\simeq 4.7$, differs significantly
for models with and without diffusion. As discussed previously diffusion causes
a decrease in helium driving, but it has little effect on driving from
hydrogen in the \ion{H}{I} ionization zone.  
The relative contribution of hydrogen to the net excitation of
a mode is larger in higher-order modes, as illustrated by
comparing the evolution of the $n=8$ and the $n=1$ modes in Fig.~\ref{fig:Aradial1.9},
and is also larger in models with diffusion because of the lower excitation from helium.

As a result, with the passage of time, the net excitation of a high-order mode is
less affected by diffusion than a low-order mode.
This can explain why, for higher-order modes shown in the
bottom panel of Fig.~\ref{fig:A1.9eta}, the growth rates are less
adversely affected by diffusion than would have been expected from a
simple extrapolation of the evolution in the young stars.

%

There are some apparent numerical effects in the work integrals 
that are caused by the manipulation of the opacity tables. First, there are
obvious spurious oscillations of the work integrals for the $n=1$ mode. Although
these oscillations change the growth rates slightly, one still gets an accurate
assessment of the differential effect of diffusion as the numerical features are the same
for all models.
Also, the apparent temperature shift towards the surface 
at 1.0 Gyr between the 1.90-ND and other models is 
probably an age effect.  The model without diffusion is 
slightly more evolved than the models with diffusion at 
a given evolution time because the former was computed using the 
Livermore Rosseland mean opacity tables rather than the 
monochromatic opacity tables. 

%

\begin{figure}[!b]
      \resizebox{\hsize}{!}{\includegraphics{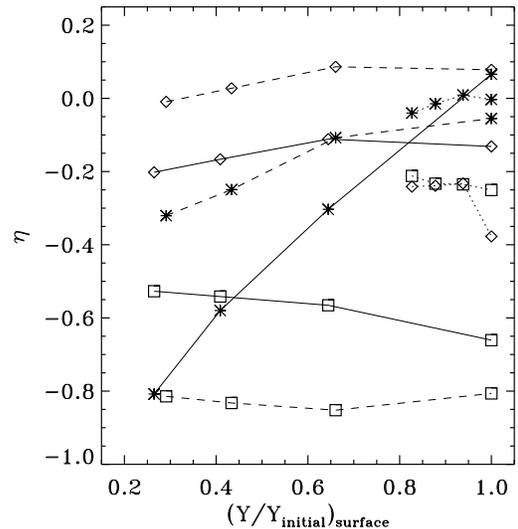}}
      \caption[]{The dependence of the growth rate $\eta$ on the superficial helium 
                 abundance is shown for four models of 1.9~M$_\odot$
                 with varying turbulence.
		 The lines link the values of $\eta$ for the
                 various modes for different models of the same age:
		 the dotted lines are for models at 30~Myr, the solid
                 line at 670~Myr and the dashed line at 1~Gyr. 
		 The higher is the turbulence in
                 a model the more to the right it will be placed.
		 Models without diffusion
                 always have $Y/Y_{\rm initial}=1.0$.
		 The models are in order of increasing
                 turbulence: 1.90R300-2, 1.90R1000-2, 1.900R10K-2 and 1.90-ND.
                 Three modes are shown: $l=0,n=1$ as diamonds,
                 $l=0,n=8$ as asterisks, and $l=1,n=-14$ (g mode) as squares.
                 }
         \label{fig:A190etaY}
   \end{figure}

Fig.~\ref{fig:A190etaY} illustrates the dependence of the growth rates on the
surface helium abundance for different ages of the 1.9~M$_{\sun}$ models.
Three modes were selected, the fundamental ($l=0,n=1$), a higher-order p mode
($l=0,n=8$) and a g mode ($l=1,n=-14$) at 100, 670 and 1000~Myr.
As one could have predicted, both p modes are very sensitive to the depletion of helium
at the surface. The higher the frequency of the mode, the larger the slope of the 
correlation.  For the g mode, there is a slight anti-correlation which shows that
it is weakly correlated to the level of driving in the iron-peak opacity bump.

The periods of the modes predicted to be overstable are summarized in Table~\ref{tab:periods1.9}.
First one can verify that the periods are within the observed period range for
$\delta$~Scuti stars. Their periods are known to range from 
30 minutes to 6 hours. The young models discussed here feature periods of 20 minutes
while older models are characterized by 
longer periods, roughly between one and four hours. While we find periods well
below the currently observed lower limit for $\delta$~Scuti stars, we know of no
reason to exclude their existence outright.  

Diffusion has little effect on the periods of oscillation.
There is a systematic trend for diffusion to increase slightly the period of a given mode.
The relative difference between the same modes in the models shown in 
Table~\ref{tab:periods1.9} is of the order of a few per~cent 
with no clear dependence on the order of the modes.
Actually, most of the difference could be consistent with only evolutionary effects
related to the use of Rosseland opacity tables for the ND model. 
If we compare the periods of the radial modes 
of models 1.90R1000-2 and 1.90R10K-2, which
are both computed with the monochromatic opacity tables,
we find period differences typically only a fifth 
of the differences between the models
calculated with differing opacities.
%

\begin{table*}
\begin{center}
\begin{tabular}{c|cccccccc}
\multispan 9 \hfill\strut Radial modes\hfill\\
\hline
\hline
age       & \multispan 8 \hfill\strut $(l,n)$ \hfill \\
(Myr)     & \strut (0,1) & (0,2) & (0,3) & (0,4) & (0,5) & (0,6) & (0,7) & (0,8) \\
\hline
\multispan 9 \hfill\strut 1.90R1000-2 \hfill \\
\hline
 1\,003   &164.30 &130.37 &104.85 & 87.32 & 75.22 &\strut &       &   \\
 1\,098   &170.70 &135.32 &108.52 & 90.39 &\strut &\strut &       &   \\
\hline
\multispan 9 \hfill\strut 1.90-ND \hfill \\
\hline
 101      &\strut &       &       &       &       &       &       & 22.38 \\
 300      &\strut &       &       &       &       &       & 28.03 & 25.14 \\
 502      &\strut &       &       &       &       & 36.11 & 32.00 &       \\
 670      &\strut &       &       & 55.39 & 47.65 & 41.66 & 37.01 &       \\
 1\,003   &163.47 &126.41 &101.49 & 84.55 & 72.54 & 63.69 & 56.66 &       \\
 1\,105   &209.26 &161.28 &128.88 &106.93 & 91.75 & 80.46 &       &       \\
\hline
\hline
\multispan 9 \strut \ \\
\multispan 9 \hfill\strut Nonradial modes\hfill\\
\hline
\hline
age       & \multispan 8 \hfill\strut $(l,n)$ \hfill \\
(Myr)     & \strut (1,1) & (1,2) & (1,3) & (1,4) & (1,5) & (1,6) & (1,7) & (1,8) \\
\hline
\multispan 9 \hfill\strut 1.90R1000-2 \hfill \\
\hline
 1\,003   & 126.46 & 101.74& 85.00 & 74.22 &       &\strut &       &   \\
 1\,098   & 105.73 & 97.90 & 87.68 &       &       &       &\strut &   \\
\hline
\multispan 9 \hfill\strut 1.90-ND \hfill \\
\hline
 101      &\strut  &       &       &       &       & 26.52 & 23.60 & 21.29 \\
 300      &\strut  &       &       &       &       & 29.77 & 26.51 &       \\
 502      &\strut  &       &       &       & 38.91 & 34.06 & 30.37 &       \\
 670      &\strut  &       & 63.04 & 52.57 & 45.03 & 39.48 &       &       \\
 1\,003   & 122.10 & 98.37 & 82.08 & 71.42 & 66.40 & 60.43 &       &       \\
 1\,105   &\strut  &       &       &       &       &       &       &       \\
\hline
\hline
\end{tabular}
\caption{Periods in minutes of overstable modes in models of a 1.9~M$_{\sun}$ star
         with and without diffusion. The upper table lists the periods of radial
         modes and the lower table that of nonradial $l=1$ p modes.
         Only the models in which at least one overstable mode was  
         found are shown.}
\label{tab:periods1.9}
\end{center}
 \end{table*}

\subsection{The effect of diffusion on the instability strip}\label{sec:AmIS}

The effect of the settling of helium on the width of the instability strip has 
already been studied by Cox et al.~(\cite{Coxetal79}). They found that the main
effect was a reduction of its width, the blue edge shifting towards
the red edge, eventually leading to the disappearance of instability when
helium was sufficiently depleted. As has been touched upon in the preceding section,
the present models seem to follow the same pattern.

In order to gain a somewhat more complete vision of the effect of diffusion
on the width of the instability strip we need to examine a larger
sample of models of A-type stars with and without diffusion.
For this purpose, additional models of 2.0 and 2.2~M$_\odot$ with various
assumed turbulence were selected. 

\begin{figure}
      \resizebox{\hsize}{!}{\includegraphics{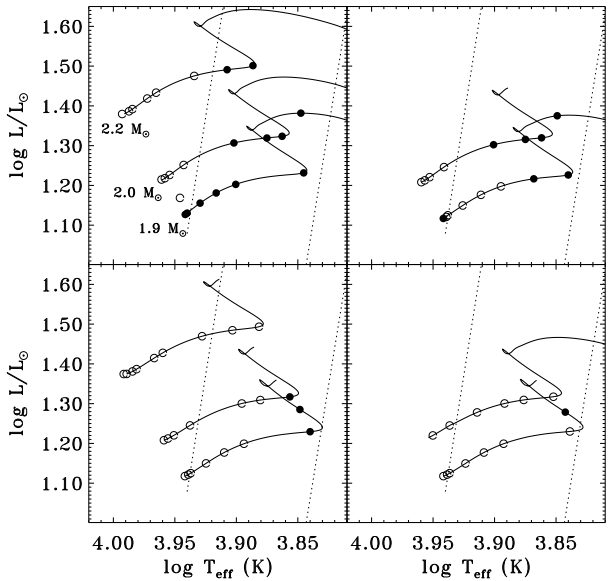}}
      \caption[]{The position of our models of 1.9, 2.0 and 2.2~M$_\odot$
                 in the HR diagram. Models for which at least one mode
                 is predicted to be overstable are marked as filled circles,
                 whereas those where we found no overstable modes
		 are marked as open
                 circles. 
		 Each of the four panels show models with similar turbulence:
                 The upper left for models without diffusion, upper right for
                 models with large turbulence (e.g. R10K-2), the lower left for models with
                 turbulence approximately representative of Am stars (R1000-2), 
                 and lower right for models with less turbulence (e.g. R300-2).}
                 
         \label{fig:AmIS}
   \end{figure}

The effect of diffusion on the width of the classical instability strip can be
estimated from Fig.~\ref{fig:AmIS} where our models which exhibit
variability are shown in the HR diagram.  
In each panel of the figure, all models have 
the same turbulence parameterization.
Clearly, the blue edge of the instability strip
is sensitive to the efficiency of the diffusion processes, 
which means that it is sensitive
to the assumed turbulence. 
It provides additional and independent constraints on the turbulence model.
The superficial metal abundances constrain the turbulence (see Richer et al.~\cite{RMT00}).
However, the superficial He abundance is poorly determined by observations. It
is much better determined via the driving of $\delta$~Scuti-type pulsations. 

In practice, each mode of oscillation has an individual blue edge and 
each mode is affected differently by diffusion. 
Tables~\ref{tab:overstable} and~\ref{tab:overstable1} and Fig.~\ref{fig:A190etaY}
suggest that the blue edge for the fundamental mode is only slightly shifted to the red
compared with normal stars 
{\sl except} if turbulence is better represented  by the R300-2 model.
The results of Richer et al.~(\cite{RMT00}) do not exclude this model,
at least not for all Am stars.
The blue edge for higher-frequency modes is shifted significantly 
as soon as the He abundance is slightly reduced.
%

As indicated in Fig.~\ref{fig:AmIS} we have considered only models hotter than
the observational red edge of the instability strip.
It is likely that the return to stability for cooler
models would be dominated by convective effects
(e.g. Houdek et al.~\cite{Houdeketal99}), which are ignored here.
On the basis of time-dependent mixing-length calculations
(Gough~\cite{Gough77}; Balmforth~\cite{Balmforth92}) one finds that the convective
effects grow rapidly as the red edge is approached
(e.g. Houdek~\cite{Houdek00}), although they have significant influence on
the stability properties also in somewhat hotter stars.
Even so, we expect that our qualitative conclusions, and
the results on differential effects of diffusion, are robust.

\subsection{Comparison with variable evolved CP stars}

The $\delta$~Delphini spectral sub-type has been shown to be a very inhomogeneous
group of stars consisting of classical Am stars, evolved Am stars and other stars
which are essentially normal A stars (Gray \& Garrison~\cite{GrayGarrison89}).
Amongst those, the $\rho$~Puppis stars are those which are thought 
to be evolved Am stars. Some of these evolved Am stars have been shown to be
variable. 
It was shown in Sect.~\ref{sec:Am} that models of evolved Am stars can be 
variable;
thus, it remains to be seen whether our evolved Am-star models
are comparable with the
observed variable CP stars. 
To verify if this is the case, the most evolved models of 1.90R1000-2,  
2.00R1000-2 and 2.20R1000-2 are compared here with the
variable metallic stars observed by Kurtz~(\cite{Kurtz76}).

\begin{figure}
      \resizebox{\hsize}{!}{\includegraphics{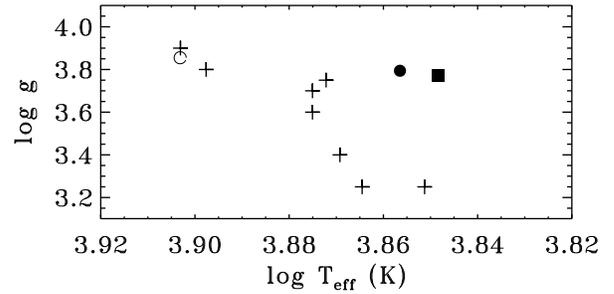}}
      \caption[]{Effective temperature vs gravity for evolved CP A stars.
                 The $\delta$~Delphini stars of 
		 Kurtz~(\cite{Kurtz76}) discussed
                 in the text are identified by the crosses. The models: filled square is for
                 1.90R1000-2 at 1094~Myr, the filled circle is for 2.00R1000-2 at 854~Myr
                 and the open circle is for 2.20R1000-2 at 670~Myr. Filled symbols represent models
                 in which overstable modes are found, the open symbol represents a model
                 in which NO overstable mode was found.
                 }
         \label{fig:dDelTg}
   \end{figure}

The positions of these stars are first shown 
in a temperature -- gravity diagram (Fig.~\ref{fig:dDelTg})
to see if the models are in an evolutionary state comparable 
with the stars in our reference sample. 
The variable 1.90R1000-2 model at 1.094~Myr and 
the variable 2.00R1000-2 model at 850~Myr are both on 
the cool side (roughly 500~K cooler than the average) of the observed stars.
Surprisingly, the 2.20R1000-2 model at 670~Myr, which is apparently a close match
to two variable metallic A stars and which is much more compatible with the
sequence of variable metallic stars than the cooler models, is predicted to be stable. 
This may or may not be a serious difficulty given the uncertainty
of effective-temperature determinations for CP stars. 
%

\begin{figure}
      \resizebox{\hsize}{!}{\includegraphics{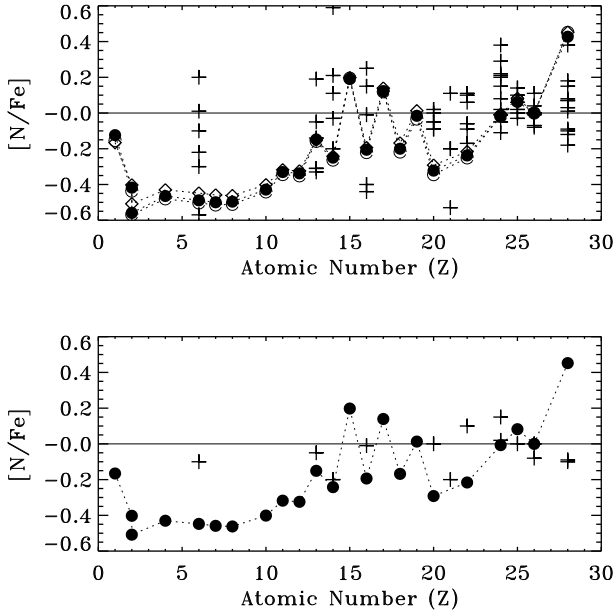}}
      \caption[]{The logarithm of the observed and predicted abundances in evolved
                 Am stars and $\delta$~Delphini stars are shown as a function of
                 the atomic number in both panels. The crosses are the observed values of 
                 Kurtz~(\cite{Kurtz76}) for $\delta$~Delphini stars. In the top
                 panel, three models are
                 shown: 1.90R1000-2 at 1094~Myr (filled circles), 
                 2.00R1000-2 at 854~Myr (open circles) and
                 2.20R1000-2 at 670~Myr (open diamonds). In the bottom panel, the predicted
                 abundances for model 2.20R1000-2 at 670~Myr are repeated along with the
                 observed abundances (Kurtz~\cite{Kurtz76}) of \object{HR1706} and \object{HR6561}.
                 }
         \label{fig:dDelabon}
   \end{figure}

The surface-abundance signatures of 
the Kurtz~(\cite{Kurtz76}) $\delta$~Delphini stars
and of the three selected models of the evolved Am stars are compared
in Fig.~\ref{fig:dDelabon}. In a second panel we repeat the
surface abundance of the evolved 2.20~M$_\odot$ but then compared with
the two stars closest to it in Fig.~\ref{fig:dDelTg}, which are
\object{HR1706} and \object{HR6561}.
These two stars have very nearly normal
abundances that are closer to the surface abundances of a 2.20R10K-2
model which would probably pulsate at that $T_{\rm eff}$ and 
$\log g$, if it behaves
similarly to the 1.9R10K-2 model.

%
Fig.~\ref{fig:dDelabon} shows that there is little variation 
in the surface-abundance profiles between the three evolved Am-star models.
This is due to the fact that the depth
of the mixed region is similar in these models since we adopted the same turbulence
parameterization in all three. 
There is a significant scatter in the observed abundances of these stars.
Nonetheless, the models lie in general within the envelope defined by the observations.
As Richer et al.~(\cite{RMT00}) pointed out, 
care must be taken when comparing with 
observed abundances due to large variations in the 
abundances determined by different observers.
Although one cannot claim that the global properties of the present models are a perfect
match to those of individual variable CP stars,
there is a qualitative agreement for the stars taken as a group. 
Better observational data for these stars would greatly help the interpretation
of the results.

\subsection{The excitation of g modes in A stars with diffusion}

As the driving regions shift deeper into the star as it evolves,
the g modes become gradually more and more excited. 
The effect of diffusion on the excitation of p modes and g modes of a
1.9~M$_{\sun}$ star was discussed in Sect.~\ref{sec:AmIS}. 
It was shown there
that p modes are stabilized through diffusion whereas g modes tend to be
excited as a result of that process. 
In the more evolved stars presented here, some g modes
are predicted to become overstable.
For example, 
we find all overstable modes in model 1.90-ND at 1.105~Gyr to be 
of the physical nature of g modes 
(the mathematical mode orders are very confusing for that model). 
However, the modes which are found to be overstable in our models are
low-order modes for which the driving from helium is significant. 
Therefore, the effect of diffusion on their excitation is similar to that 
of low-order p modes, i.e.,
the settling of helium leads to a reduction of their excitation.

As an example for one of these g modes, 
we get the following growth parameters for
the $l=1,n=-1$ mode in the different models of 1.9~M$_{\sun}$ at roughly 1~Gyr, 
in order of decreasing helium
content in the driving region: 0.0864 for 1.9-ND, 
0.0891 for 1.9R10K-2, 0.0294 for 1.9R1000-2 and $-0.0098$ for 1.9R300-2.

The detection of g modes in $\delta$~Scuti stars has been problematic.
Models for evolved A stars predict a very dense spectrum of overstable g modes
while observations so far show little evidence of such variations (Guzik~\cite{Guzik00}).
In addition, Breger \& Beichbuchner~(\cite{BregerBeichbuchner96}) show that there is no observational
evidence for long-period g modes (of the nature of those observed in cooler
$\gamma$~Doradus stars) in $\delta$~Scuti stars. The models suggest that they would not
be easier to find in evolved Am stars, especially considering that convective effects
should become increasingly important in models leaving the main sequence.

\section{Is there a seismic signature of diffusion in $\delta$~Scuti stars?}
\label{sec:signature}

In general,
$\delta$~Scuti stars are fairly fast rotators in which diffusion is not
expected to be important. In those stars for which diffusion could be important
it is legitimate to ask in what measure it could influence their
modeling.

The high-amplitude $\delta$~Scuti (HADS) stars are typically slow rotators and
as such are candidates to become Am stars. 
They are characterized by high-amplitude
pulsations compared with their normal counterparts. All evidence currently 
points to HADS being normal stars (H{\o}g \& Petersen~\cite{HogPetersen97}) 
following normal stellar evolution (Petersen \& Christensen-Dalsgaard~\cite{PetersenJCD96}).
They are not distributed uniformly in the instability strip but occupy only
a strip some 300\,K wide.
If the high amplitudes were linked to a particular non-linear effect caused
by the low rotational velocity this could have
interesting repercussions on the stability of other slowly rotating A stars, such as the Am stars.
Certainly, nothing in our results would lead us to believe that diffusion, by itself, would
generate higher-amplitude modes in A stars.
The lack of abundance anomalies in HADS might simply be due to their 
advanced evolutionary state and the depth of the convection zone 
at that time.

As has been briefly touched upon in the introduction (Sect.~\ref{sec:obs})
some $\delta$~Scuti stars may exhibit abundance anomalies.
These anomalies are not small contrary to what one might expect. In fact, 
one can see a scatter of [Ca/Fe] from $-0.6$  to $0.7$ 
and depletions of carbon of up to [C/Fe] $\simeq -1.3$ 
in Russell's~(\cite{Russell95}) results.
Referring to Fig.~\ref{fig:dDelabon} it is evident
that the anomalies reported by Russell are higher than those 
observed in evolved Am stars. 
Similar results have been also published by Rachkovskaya~(\cite{Rachkovskaya94}) 
and her other work referenced therein.

Although the anomalies have the overall appearance of diffusion,
{\sl e.g.} depleted C and Ca, normal
Si and enhanced Fe and Ti, the scatter of the observed abundances is so 
large as to make any correlation between the abundances of the 
different elements very difficult.
There is no systematic signature of diffusion as found in Am stars
(cf. Richer et al.~\cite{RMT00})
or even in the more evolved Am stars shown in Fig.~\ref{fig:dDelabon}. Moreover,
some of those stars exhibit a $v\sin i$ as large as 100 to 150\,km\,s$^{-1}$.

In other more standard $\delta$~Scuti stars the abundance anomalies 
are expected to be small. In such cases, the effect of diffusion 
on the seismology of those stars
would be subtle. Previous experience in lower-mass stars 
(Turcotte \& Christensen-Dalsgaard~\cite{TJCD98}) suggests that any seismic
signature of diffusion would be overwhelmed by other effects 
such as convective-core overshoot, for example. 
In the event that the iron convection zone would develop, one could
expect a clear seismic signature if a sufficient number of oscillation modes were identified.
Guzik~(\cite{Guzik93}) pointed out that helium settling
may have some consequence on
observed properties of $\delta$~Scuti stars, such as the light curve.

\section{Pulsation and turbulence} \label{discussion_A}

In their discussion of metallicism and pulsations, Kurtz and collaborators have
made the point that pulsations involve rapid motion to a 
substantial depth in the star. 
They argue that velocity of the displacement generated by the pulsation
remains high throughout the region in which the abundance anomalies
are thought to be formed (referring for example to the eigenfunctions
displayed in Fig. 8.2 of 
Cox~\cite{Cox80}). They speculate that the high velocities might
generate turbulent mixing and therefore inhibit the occurrence of
metallicism.

Although the velocity of the radial displacement might be large, turbulence
is expected to be generated not by the speed of the displacement itself but rather by 
velocity gradients: a fast but uniform displacement will not 
become turbulent. The scale on which the displacement occurs is
much greater than other relevant scales, e.g. the pressure scale height.
The gradient of the displacement velocity is
small over those scales and one would not expect turbulence to be 
necessarily generated as a result.

Our understanding of the extent to which pulsations are laminar
or generate turbulence is too incomplete 
for us to speculate further on the link
between the turbulence assumed in Am stars and the observed pulsations.

\section{Conclusion} \label{sec:conclusion}

We have investigated the impact of diffusion on the stability of A-type 
main-sequence Population~I stars. The models include the consistent abundance
evolution of all important elements and its impact on stellar structure. 
In these models helium remains present in the \ion{He}{II} ionization zone
and the opacity in the iron opacity bump increases substantially, raising
the possibility of a strong relationship between variability and diffusion.

We present evidence that young Am stars are stable against driving from
the $\kappa$ mechanism and that, as the stars evolve, they become
unstable, but only when near the red edge of the instability strip.
Hot Am stars need to be more evolved than cool Am stars before variability
can occur.
The blue edge of the instability strip for metallic A stars is sensitive to the
magnitude of the abundance variations and is thus indicative of the
depth of mixing by turbulence. 

The stability of A stars is more sensitive to the evolution of the abundance of
helium than to the accumulation of iron-peak elements.
In stars with very little turbulence the iron-peak driving region can become
convectively unstable thereby reducing the radiative flux there and negating the
driving effect of the enhanced opacity bump. However, in the models which 
are representative of Am stars, the turbulence is high enough to prevent the formation
of that iron convection zone while allowing a significant increase of the 
opacity bump due to iron-peak elements.  Still,  only a marginal positive effect
can be seen in the long-period g modes. The higher-frequency modes depend mostly
on the helium abundance and somewhat on the hydrogen abundance.

There are a number of caveats relevant to the present work.

There is no direct link between the normalized growth rates $\eta$ and the
actual amplitude of the pulsations as evidenced by the lower number of 
modes observed in $\delta$~Scuti stars relative to their predicted number. 
Additionally, comparisons to models at solar composition 
by J.~Christensen-Dalsgaard and W.~Dziembowski 
have shown that the growth rates are sensitive to the details of the modeling. In general, in the
stars we compared at 1.8 and 1.9~M$_{\sun}$, the growth rates were lower in our models
than in either of their models.

The treatment of convection in the present work is simplistic. 
Convection and turbulence
are known to damp pulsation to a certain extent. 
The most striking and well known evidence of the effect of convective 
damping is the red edge of the instability strip 
[Gough~(\cite{Gough77});
Gonczi \& Osaki (1980);
Balmforth \& Gough~(\cite{BalmforthGough88}); 
see also Buchler et al.~(\cite{BYKG99}) for a recent review].
Turbulence affects pulsations in two ways: 1) through turbulent viscosity, which always
dampens pulsations, and 2) through the phase difference between entropy variations
and the modulation of the convective flux, which can either excite or dampen pulsations.
It is the latter effect which is responsible for the red edge of the classical instability
strip. While the interaction between turbulent convection and pulsations has been studied,
the effect of turbulence outside of the convection zones is unknown.

In the models including diffusion, the coupling between turbulence
and diffusion might be very important. 
For the present models reproducing Am stars, the turbulence extends to a significant
depth below the convection zone. While the energy flux  related 
to the turbulence is expected to be very small, the effect of turbulent 
viscosity might not be negligible.  For stars in which the iron convection
zone is allowed to form, three separate convection zones exist
with highly turbulent regions between them.
The behavior of pulsations in such stars might be heavily 
affected by a proper treatment of convective effects.

While it is true that our analysis does not 
take this into account, one should not forget that in the standard models
without diffusion an {\sl ad hoc} mixing mechanism is implied to prevent the
formation of abundance gradients. This presumably turbulent mixing
(at least in the more slowly rotating $\delta$~Scuti stars) is also 
never taken into account. It is, by hypothesis, larger than that in the models
where atomic diffusion is important.
Under these circumstances, 
we may perhaps consider the simple treatment of convection and
turbulence applied here as adequate for identifying
the differential effect of diffusion on stellar stability.

The NMM used here still contain an element of arbitrariness in that they extend
turbulent mixing a little beyond that expected from iron convection zones, without
providing a physical mechanism for this extension. This extension is required to
fit the observed surface abundances of Am stars.
The extension is, however,
sufficient to cause the iron-peak abundances to 
decrease sufficiently for the convection zone to disappear in models
representative of these stars.
The models do not include the potential 
effects of mass loss or rotation, although differential
rotation is one mechanism that is now being investigated as a source of the
instabilities that could provide this mixing zone extension.

When compared against each other the present models do, however, 
illustrate the effect of diffusion on the stability of main-sequence A
stars as well as present modeling allows.

\section*{ACKNOWLEDGEMENTS}
This work was supported in part by the Danish National
Research Foundation through its establishment of the Theoretical
Astrophysics Center. We are deeply indebted to Wojtek Dziembowski
for his pulsation stability code and to G\"unter Houdek for his help in
implementing his interpolation routines. We also acknowledge the 
very valuable comments of W. Dziembowski on the computations, as well
as on an earlier version of the manuscript. Thanks are extended to
D. O. Gough, D. Kurtz and J.-P. Zahn for insightful discussions,
and J. O. Petersen for careful reading of the originally submitted manuscript.

\end{document}